\documentclass[final,5p,times,twocolumn]{elsarticle}

\usepackage[utf8]{inputenc} 
\usepackage[T1]{fontenc}
\usepackage{dcolumn}
\usepackage{braket}
\usepackage{float}
\usepackage{graphicx}
\usepackage[colorlinks=true,linkcolor=blue,citecolor=blue,urlcolor=blue]{hyperref}
\usepackage{orcidlink}
\usepackage{amsmath}
\usepackage{amssymb}
\usepackage{lipsum}
\usepackage{textcase}
\usepackage{subcaption}
\usepackage[normalem]{ulem}
\usepackage{bm}
\usepackage{color}
\usepackage{amsfonts}
\usepackage{slashed}
\usepackage{pstricks}
\usepackage{multirow}
\usepackage{hyperref}
\usepackage{array}
\allowdisplaybreaks
\usepackage{epsf}
\usepackage[nice]{nicefrac}
\usepackage{placeins}

\biboptions{sort&compress}
\begin{document}

\begin{frontmatter}

\title{Impact of \textit{ab initio} E1 strength 
on neutron capture rates along the nickel isotopic chain}

\author[first,second]{Francesca~Bonaiti\orcidlink{0000-0002-3926-1609}}
\ead{bonaiti@frib.msu.edu}
\affiliation[first]{organization={Facility for Rare Isotope Beams, Michigan State University}, 
            city={East Lansing},
            postcode={48824}, 
            state={Michigan},
            country={USA}}
\affiliation[second]{organization={Physics Division, Oak Ridge National Laboratory}, 
            city={Oak Ridge},
            postcode={37931}, 
            state={Tennessee},
            country={USA}}
\author[third]{Joanna E. Sobczyk\orcidlink{0000-0003-4698-9339}}
\ead{joanna.sobczyk@chalmers.se}
\affiliation[third]{organization={Department of Physics, Chalmers University of Technology},
            city={Göteborg},
            postcode={SE-412 96}, 
            country={Sweden}}

\begin{abstract}
Nucleosynthesis simulations of the $r$-process require neutron capture rates for neutron-rich nuclei, many of which lie beyond experimental reach. Predicting these rates relies on theoretical calculations of the low-energy electric dipole (E1) strength near the neutron threshold.
We present the first \emph{ab initio} computation of the E1 strength along the nickel isotopic chain, for $^{58}$Ni and the even-even isotopes $^{68-80}$Ni. The E1 strength is obtained with the Lorentz integral transform coupled-cluster (LIT-CC) method, extended to axially symmetric reference states to reach open-shell nuclei, using two- and three-nucleon interactions from chiral effective field theory. The approach is benchmarked against the measured photoabsorption cross section of $^{58}$Ni and the E1 strength of $^{68}$Ni. Along the chain we find low-lying E1 strength above the neutron threshold that grows with neutron number. Including it in the reaction code TALYS enhances the neutron capture rates forming $^{68-78}$Ni between 3 and 10~GK, with an effect that increases with temperature and neutron number and an uncertainty dominated by the inversion of the integral transform. For $^{80}$Ni, where strong pygmy dipole strength lies just above threshold, the enhancement sets in already around 0.5~GK. Together, these results open the way to \textit{ab initio} input for neutron capture rates along isotopic chains far from stability.
\end{abstract}


\end{frontmatter}

\section{Introduction}
\label{intro}
Understanding how the elements heavier than iron are formed is one of the overarching questions in nuclear astrophysics. About half of them are produced in the rapid neutron-capture process ($r$-process)~\cite{burbidge1957,seeger1965}, which takes place in extremely neutron-rich astrophysical environments~\cite{cote2018} and proceeds through a sequence of $\beta$-decays and neutron captures far from stability. The final abundance pattern is sensitive to individual neutron capture rates, particularly during the $r$-process freeze-out~\cite{beun2008,surman2009,arcones2011,mumpower2012,kedia2026}. Direct measurements of neutron capture cross sections on short-lived nuclei are experimentally challenging, while indirect approaches face increasingly limited statistics toward the neutron-rich extremes~\cite{escher2012,ratkiewicz2019,larsen2019}. Theoretical predictions are therefore essential to cover the regions of the nuclear chart far from stability~\cite{mumpower2016}.

The established approach for computing neutron capture rates relies on the Hauser-Feshbach statistical model~\cite{hauser1952}, based on the compound-nucleus picture~\cite{bohr1939}, using codes such as TALYS~\cite{koning2023}, CoH$_3$~\cite{kawano2021} and YAHFC~\cite{ormand2021}. These calculations require a set of nuclear inputs, the most important being the $\gamma$-ray strength function and the nuclear level density. The neutron–nucleus optical potential, which also enters the calculation, was found to contribute less to the reaction rate uncertainty~\cite{nikas2020}, although its impact may increase farther from stability.

Under the Brink-Axel hypothesis~\cite{brink1955,axel1971,bartholomew1973}, the $\gamma$-ray strength function for transitions from the excited states of the compound nucleus can be determined from the photoabsorption cross section of the ground state. The latter is dominated by electric dipole (E1) strength, concentrated in the giant dipole resonance (GDR), while magnetic dipole (M1) transitions have been found to contribute at low $\gamma$-ray energies~\cite{larsen2010,mumpower2017,larsen2018,sasaki2023}, in particular below the neutron separation energy, as it was observed in several nuclei~\cite{schwenger2013,brown2014,liddick2016,simon2016,larsen2019,guttormsen2022,spyrou2024,ronning2026}.
In neutron-rich nuclei the E1 strength can exhibit additional strength in the vicinity of the separation energy~\cite{klimkiewicz2007,rossi2013,aumann2013}, often associated with the pygmy dipole resonance (PDR)~\cite{inakura2011,rocamaza2012}. Several studies have shown that this low-lying E1 strength can significantly affect neutron capture rates~\cite{goriely1998,goriely2002,litvinova2009,tsoneva2015}.
Different theoretical frameworks have been employed to describe it, from phenomenological Lorentzian models~\cite{kopecky1990,goriely2019} to approaches based on energy density functionals (EDF), such as the quasiparticle random-phase approximation and its extensions~\cite{litvinova2009,martini2016,goriely2018,xu2021,kaur2025,ghosh2026}, and the shell model~\cite{brown2014,larsen2018,sieja2017,schwengner2017,gorton2026a,gorton2026b}.

In this work we address this problem for the first time from first principles, computing the E1 response (i.e. strength) of neutron-rich nickel isotopes with an \textit{ab initio} many-body approach~\cite{hergert2020,ekstrom2022}, which starts from nuclear interactions derived from chiral effective field theory~\cite{weinberg1990,epelbaum2009,machleidt2011} and solves the many-body problem with systematically improvable methods~\cite{hagen2014,dickhoff2004,lu2019,hergert2016,stroberg2019}. Recent advances have extended \textit{ab initio} calculations to electromagnetic responses of medium-mass nuclei~\cite{bacca2026,raimondi2019,porro2024,burrows2025,bonaiti2026_tdcc,bonaiti2026_monopole,porro2026}, paving the way for independent constraints on the E1 response of neutron-rich nuclei that can guide extrapolations of phenomenological and EDF-based models. In particular, the Lorentz integral transform coupled-cluster (LIT-CC) method~\cite{bacca2013,bacca2014} provides access to response functions without explicitly constructing scattering states. It has been applied to closed-shell nuclei~\cite{simonis2019,kaufmann2020,sobczyk2021,sobczyk2024,acharya2025,lehr2026} and extended to selected open-shell nuclei in the vicinity of shell closures~\cite{bonaiti2024,brandherm2025,marino2025}. However, it was pointed out that these frameworks lack sufficient particle-hole correlations for a reliable description of the electric dipole response. To study the E1 response along isotopic chains, we extend here the LIT-CC method to an axially symmetric formulation that allows rotational symmetry to be broken in the reference state.

We focus on the nickel isotopic chain up to $^{80}$Ni. It is astrophysically relevant, as neutron capture rates on neutron-rich nickel isotopes influence the abundances below $A\approx 100$ in the weak $r$-process~\cite{surman2014,mumpower2024}, and it allows for comparisons with data for $^{58}$Ni and $^{68}$Ni, with further constraints expected from ongoing measurements at FRIB~\cite{spyrou_priv}. We concentrate on the E1 response above the first particle emission threshold, which carries the strength relevant for neutron capture.

The article is organized in the following way. In Section~\ref{formalism}, we present the computational methods employed in this work. In Section~\ref{validation}, we benchmark the new LIT-CC extension against data for $^{58}$Ni, we test our predictions in the neutron-rich region using data for $^{68}$Ni and we discuss our results for the E1 response of neutron-rich even-even Ni isotopes, from $^{70}$Ni to $^{80}$Ni. We assess uncertainties associated with model-space convergence, the inversion of the Lorentz integral transform, breaking of rotational invariance and, for $^{68,78}$Ni, the Hamiltonian dependence. In Section~\ref{astro_rates}, we use the resulting E1 strength as input to Hauser-Feshbach calculations with the reaction code TALYS to assess the impact of low-lying E1 strength on neutron capture rates. Finally, we draw our conclusions in Section~\ref{conclusions}.

\section{Formalism}
\label{formalism}
\subsection{Hauser-Feshbach theory for neutron capture reactions}
The quantity of interest for nucleosynthesis simulations is the astrophysical reaction rate $N_A\braket{\sigma v}$ at a given temperature $T$. The latter is defined as~\cite{rauscher2000}
\begin{equation}
    N_A\braket{\sigma v} = N_A \left(\frac{8}{\pi\mu}\right)^{1/2} \frac{1}{(kT)^{3/2}} \int_0^{\infty} \sigma(E) E \;\mathrm\exp\left({-\frac{E}{kT}}\right) dE
\end{equation}
where $N_A$ is Avogadro's number, $\mu$ is the reduced mass and $\sigma(E)$ is the cross section of the process of interest. 

We compute the neutron capture cross section $\sigma_{n\gamma}(E)$ in the framework of Hauser-Feshbach theory, that we briefly review below following Refs.~\cite{rauscher2000,mumpower2017,larsen2019}. According to the latter, neutron capture reactions are assumed to happen in two steps: (i) a compound nucleus is created; (ii) the compound nucleus undergoes $\gamma$-ray decay. These two stages are assumed to be completely independent from each other: the compound nucleus loses memory of how it was formed and decays are dominated by its statistical properties. 

Under these assumptions, for a target nucleus with spin $J_t$ hit by an incoming neutron with energy $E_n$, the neutron capture cross section can be computed as
\begin{equation}
    \sigma_{n\gamma}(E_n) = \frac{\pi}{k_n^2} \frac{1}{(2J_t + 1)(2J_n + 1)} \sum_{J\pi} (2J+1) \frac{T_n T_{\gamma}}{T_n + T_{\gamma}}
\end{equation}
where $T_n$ is the neutron transmission coefficient, $T_{\gamma}$ is the $\gamma$-ray transmission coefficient, $k_n$ is the wave number of the incident neutron and the sum runs over the all possible spin $J$ and parity $\pi$ states of the compound nucleus. To account for the fact that the decay of the state is not fully statistical, width fluctuation corrections are typically introduced~\cite{hilaire2003,kawano2015}. 

The neutron transmission coefficient $T_n$ is computed by solving the Schrödinger equation with an optical potential for the neutron-nucleus interaction. In this work, we will focus on the $\gamma$-ray transmission coefficient $T_{\gamma}$ defined as
\begin{equation}
    T_{\gamma} = \sum_{J\; XL} \int_0^{E'} 2\pi E_{\gamma}^{2L + 1} f_{XL}(E_{\gamma}) \rho(E_x, J) dE_x
\end{equation}
where $E' = S_n + E_n$ is the total excitation energy of the compound nucleus, $E_x$ is the excitation energy of the final state, $S_n$ is the neutron separation energy, $E_{\gamma} = E' - E_x$ is the emitted $\gamma$-ray energy, $\rho(E_x, J)$ is the level density of the compound nucleus and $f_{XL}(E_{\gamma})$ is the $\gamma$-ray strength function of multipolarity $L$ and type $X= E, M$ for electric or magnetic transitions, respectively. 

In this work, we study the sensitivity of neutron capture rates to electric dipole (E1) transitions, which together with M1 dominate $T_\gamma$. In this case, the $\gamma$-ray strength function is defined as
\begin{equation}
    f_{\mathrm E1}(E_{\gamma}) = \frac{\sigma_{\gamma}(E_{\gamma})}{3(\pi \hbar c)^2 E_{\gamma}}  
    \label{gamma_sf}
\end{equation}
where $\sigma_{\gamma}(E_{\gamma})$ is the photoabsorption cross section. The latter is computed as 
\begin{equation}
    \sigma_{\gamma}(E_{\gamma}) = 4\pi^2 \alpha E_{\gamma} R(E_{\gamma})
    \label{photoabs_xs}
\end{equation}
where $R(E_{\gamma})$ is the nuclear E1 response function
\begin{equation}
   R(E_{\gamma}) = \sum_{\mu} |\braket{\Psi_{\mu}|\hat{\Theta}|\Psi_0}|^2 \delta(E_{\mu} - E_0 - E_{\gamma})\, .
   \label{response}
\end{equation}
$\ket{\Psi_0}$ is the ground state of the nucleus, and $\ket{\Psi_{\mu}}$ are the excited states with energy $E_{\mu}$ that can be reached from the ground state via the transition operator $\hat\Theta$ (corresponding to the electric dipole operator in this case). In the following, we will present the computational framework for this observable within coupled-cluster theory. 

\subsection{Coupled-cluster theory for ground and excited states}
The starting point of our many-body computations is the nuclear intrinsic Hamiltonian
\begin{equation}
\label{H}
H = \frac{1}{2mA} \sum_{i<j}^A (\vec{p}_i -\vec{p}_j)^2 + \sum _{i<j}^A V_{ij} + \sum_{i<j<k}^A W_{ijk}  \,,  
\end{equation}
where $m$ is the nucleon mass, $A$ is the mass number, $V_{ij}$ is the two-body force and $W_{ijk}$ is the three-body force. In this work, we employ the $\Delta$NNLO$_{\rm GO}$(394) interaction~\cite{jiang2020}, which is formulated at next-to-next-to-leading-order in the chiral expansion and includes the $\Delta$ isobar as a degree of freedom. To assess the dependence of our results on the choice of Hamiltonian, in the case of $^{68,78}$Ni we also perform computations with the newly developed N$^3$LO$_{\rm Texas}$~\cite{hu2026} interaction. The latter is a $\Delta$-less potential, including two- and three-body forces at fourth and third chiral order, respectively. 

We solve the Schr\"odinger equation for $H$ with coupled-cluster (CC) theory~\cite{coester1958,coester1960,cizek1966,kuemmel1978,bartlett2007,hagen2014}, starting from a Hartree-Fock reference state $\ket{\Phi_0}$. For closed-shell nuclei the reference is rotationally invariant and carries total angular momentum $J=0$. For open-shell nuclei, following Refs.~\cite{novario2020,hagen2022}, the Hartree-Fock calculation is allowed to break rotational invariance while conserving axial, parity and time-reversal symmetry, so that the projection $j_z$ of the angular momentum on the symmetry axis is conserved. The reference is built by filling the partially occupied shells at the Fermi surface from low to high $|j_z|$, which leads to a prolate $\ket{\Phi_0}$. The underlying harmonic oscillator basis is characterized by the maximum number of shells $N_{\rm max}$ and the frequency $\hbar\Omega$, which control model-space convergence. 

Correlations are introduced through the exponential ansatz $e^T\ket{\Phi_0}$, where the cluster operator $T$ is a sum of particle-hole (p-h) excitations. The ground-state problem becomes an eigenvalue problem for the similarity-transformed Hamiltonian $\overline{H}_N = e^{-T}H_Ne^{T}$,
\begin{equation}
    \overline{H}_N \ket{\Phi_0} = E_0 \ket{\Phi_0}\,,
    \label{se-cc}
\end{equation}
with $H_N$ normal-ordered with respect to $\ket{\Phi_0}$. In this work we truncate $T$ at the 2p-2h level, i.e. the CC singles and doubles (CCSD) approximation.

Excited states are obtained with the equation-of-motion CC (EOM-CC) method~\cite{stanton1993,bartlett2003,bartlett2007}, in which right and left eigenstates of $\overline{H}_N$ are written as
\begin{equation}
    \ket{\Psi_{\mu}} = R_{\mu} e^T \ket{\Phi_0}\,, \qquad
    \bra{\Psi_{\mu}} = \bra{\Phi_0} L_{\mu}  e^{-T}\,,
\label{ansatz}
\end{equation}
with $R_\mu$ and $L_\mu$ linear (de-)excitation operators, truncated at the 2p-2h level consistently with the ground state. Since $\overline{H}_N$ is non-Hermitian, both are needed to compute transition strengths. $L_0$ denotes the de-excitation operator of the left ground state.

\subsection{Lorentz Integral Transform technique}
\label{sec:lit}
In the CC formalism the response function of Eq.~(\ref{response}) becomes
\begin{equation}
\begin{split}
    R(E_{\gamma}) = &\sum_{\mu} \braket{\Phi_0|L_0 \overline{\Theta^{\dagger}} R_{\mu}|\Phi_0} \\ &\times \braket{\Phi_0|L_{\mu}\overline{\Theta}|\Phi_0} \delta(E_{\mu} - E_0 - E_{\gamma}), 
    \label{response-cc}
\end{split}
\end{equation}
where $\overline{\Theta} = e^{-T} \hat\Theta e^T$ is the similarity-transformed transition operator. Above the first particle threshold the sum over $\mu$ runs over continuum states, whose explicit construction would require all breakup channels of the nucleus. The Lorentz integral transform (LIT) technique~\cite{efros1994,efros2007} avoids this: in the LIT-CC method~\cite{bacca2013,bacca2014}, the transform
\begin{equation}
    \mathcal{L}(\sigma, \Gamma) = \frac{\Gamma}{\pi}\int dE_{\gamma} \frac{R(E_{\gamma})}{(E_{\gamma}-\sigma)^2 + \Gamma^2}
\end{equation}
is obtained as $\mathcal{L}(\sigma, \Gamma) = \braket{\tilde{\Psi}_L(\sigma)|\tilde{\Psi}_R(\sigma)}$ from the solutions of the EOM-CC equations with a source term
\begin{align}
    (\overline{H}_N - E_0 - \sigma) \ket{\tilde{\Psi}_R(\sigma)} &= \overline{\Theta} \ket{\Phi_0}\,, \\
    \bra{\tilde{\Psi}_L(\sigma)} (\overline{H}_N - E_0 - \sigma) &= \bra{\Phi_0} L_0 \overline{\Theta}\,,
    \label{lit_cc_equations}
\end{align}
which have bound-state-like solutions and can be solved with the Lanczos algorithm~\cite{bacca2014}. For open-shell nuclei, where the axially symmetric basis makes the calculations memory-intensive in large model spaces, we instead compute the LIT from Chebyshev moments, which avoids storing the Lanczos vectors~\cite{sobczyk2022}. The response is then recovered from $\mathcal{L}(\sigma,\Gamma)$ by numerical inversion (see Sec.~\ref{inversion}).

To reach open-shell nuclei, we have implemented the EOM-CC and LIT-CC in an axially symmetric single-particle basis, where $j_z$ rather than $J$ is a good quantum number. The E1 operator then has to be treated component by component, and we compute the LIT separately for its three spherical components $\kappa=0,\pm1$ 
\begin{equation}
    \hat{D}_{\kappa} = \sqrt{\frac{4\pi}{3}}\sum_{i=1}^A r_i Y^1_{\kappa}(\hat{r}_i)\left(\frac{1+\tau^z_i}{2}\right)\,.
    \label{eq:dipole_op}
\end{equation}
where $\tau^z_i$ is the $z$-axis projection of the $i$-th nucleon's isospin. By time-reversal invariance, the $\kappa=\pm1$ results coincide. Restoring rotational invariance would require angular momentum projection, which we leave to future work. Instead, we use the difference between the $\kappa=0$ and $\kappa=1$ results, which would vanish in a rotationally invariant calculation, as an estimate of symmetry-breaking effects, and, in analogy to previous works (see e.g. Ref.~\cite{kanadaenyo2005}), approximate the projected result by the average over the three components,
\begin{equation}
    R(E_{\gamma}) = \frac{1}{3}\sum_{\kappa=0,\pm1} R_{\kappa}(E_{\gamma}) = \frac{1}{3} R_{\kappa = 0}(E_{\gamma}) + \frac{2}{3} R_{\kappa = 1}(E_{\gamma}).
    \label{eq:average}
\end{equation}

\subsection{Inversion of the integral transform}
\label{inversion}
The response function is recovered from $\mathcal{L}(\sigma,\Gamma)$ by a numerical inversion. We follow the well-established protocol of Refs.~\cite{efros2007,andreasi2005}
and expand the response on a small set of basis functions,
\begin{equation}
    R(E_\gamma) = \sum_{n=1}^{N} c_n g_n(E_\gamma)\,,
    \label{eq:expansion}
\end{equation}
chosen to encode the known physical behaviour of the response: at the threshold $E_{\rm thr}$ of the first two-body breakup channel the strength rises as $(E_\gamma-E_{\rm thr})^{l+1/2}$, with $l$ the relative orbital angular momentum of the emitted nucleon, while at high energies it decays exponentially. Several families of such functions have been proposed in the past~\cite{efros2007,andreasi2005}; they typically depend on two or three nonlinear parameters, which are optimized in a least-squares fit. Here we adopt the ansatz
\begin{equation}
    g_n(E_\gamma) = x^{3/2}\,\frac{e^{-x/(\beta n)}}{1+e^{-(x-\delta)}}\,, \qquad x = E_\gamma - E_{\rm thr} > 0\,,
    \label{eq:basis}
\end{equation}
and $g_n(E_\gamma)=0$ below threshold. The factor $x^{3/2}$ corresponds to $l=1$ nucleon emission, while the Fermi-type factor in the denominator adds flexibility in the low-energy region.
We found this choice to be robust along the whole isotopic chain considered here, reproducing both the low-lying pygmy strength and the giant dipole resonance (GDR).

The threshold energy $E_{\rm thr}$ is the neutron separation energy $S_n$ for $^{68-80}$Ni and the proton separation energy $S_p$ for $^{58}$Ni, where proton emission is the first open channel. We use the separation energies from the Atomic Mass Evaluation 2020~\cite{ame2020}, 
which for the closed-shell $^{68}$Ni and $^{78}$Ni lie $0.3$--$0.5$~MeV above those obtained with the $\Delta$NNLO$_{\rm GO}(394)$ and the N$^3$LO$_{\rm Texas}$ Hamiltonians. The only exception is $^{68}$Ni, whose coupled-cluster discretized spectrum contains a state just above the computed threshold and below the experimental one. In this case we use the theoretical $S_n$ for inversion and afterwards shift the resulting response so that its threshold coincides with the experimental value. For $^{74,76,78,80}$Ni, where no measured values exist, we use the recommended ones~\cite{ame2020}.

The expansion of Eq.~(\ref{eq:expansion}) describes only the continuum part of the response. For the closed-shell $^{68}$Ni and $^{78}$Ni, where the bound part of the EOM-CC spectrum is directly accessible, we find in each case a single low-lying $1^-$ state at about 1~MeV, 
carrying a tiny fraction of the E1 strength.
Its contribution is negligible on the scale of the inversion uncertainty.
A single bound dipole state of this size is irrelevant for the neutron capture rates discussed in Sec.~\ref{astro_rates}, and we have not searched for bound states in the open-shell isotopes since their accurate description in the symmetry-broken framework would require angular-momentum projection, which we leave to future work. In the following we therefore concentrate on the E1 strength above the particle threshold.

For given values of $\beta$, $\delta$ and $N$ in Eq.~\eqref{eq:expansion}, the expansion coefficients $c_n$ are obtained by minimizing $\|\mathcal{L}(\sigma,\Gamma) - \mathcal{L}_{\rm rec}(\sigma,\Gamma)\|$ over a grid of $\sigma$ values at fixed $\Gamma$, 
where
\begin{equation}
    \mathcal{L}_{\rm rec}(\sigma,\Gamma) = \frac{\Gamma}{\pi}\sum_{n=1}^{N} c_n \int dE_\gamma\, \frac{g_n(E_\gamma)}{(E_\gamma-\sigma)^2+\Gamma^2}
    \label{eq:lit_rec}
\end{equation}
is the transform of the trial response. The nonlinear parameters $\beta$ and $\delta$ are then varied over a range of values, 
together with the number of basis functions, $N = 11$--$14$. All resulting solutions are retained, and the envelope of this family defines the inversion uncertainty band shown in our results. We deliberately use the same inversion setup for all isotopes, and for two values of $\Gamma=5,\, 10$ MeV. This conservative choice leads to a large family of admissible solutions, and hence to a sizable inversion uncertainty, which could be reduced in the future by a stricter acceptance criterion or by a different inversion strategy.

To assess the quality of the inversions we need a scale against which to judge the residual $\mathcal{L}-\mathcal{L}_{\rm rec}$. Since our open-shell results are not fully converged in the largest accessible model space, we take as this scale the many-body uncertainty of the transform itself. We estimate it from $^{78}$Ni, for which we can reach $N_{\rm max}=14$, by comparing the transforms $\mathcal{L}_i$ obtained for $N_{\rm max}=12,14$ and $\hbar\Omega=10,12,16$~MeV with the best-converged one, $\mathcal{L}_{\rm ref}\equiv\mathcal{L}_{N_{\rm max}=14}^{\hbar\Omega=10}$,
\begin{equation}
    \epsilon(\sigma) = \max_i \frac{|\mathcal{L}_{\rm ref}(\sigma) - \mathcal{L}_i(\sigma)|}{\mathcal{L}_{\rm ref}(\sigma)}\,.
    \label{eq:epsilon}
\end{equation}
For each inversion we then compute a ``reduced $\chi^2$''
\begin{equation}
    \chi^2 = \frac{1}{N_\sigma}\sum_{i=1}^{N_\sigma} \left[\frac{\mathcal{L}(\sigma_i)-\mathcal{L}_{\rm rec}(\sigma_i)}{\epsilon(\sigma_i)\,\mathcal{L}(\sigma_i)}\right]^2 ,
    \label{eq:chi2}
\end{equation}
where the sum runs over the $N_\sigma$ grid points. This quantity does not have a strict statistical interpretation, but it provides a handle on the accuracy of a given solution. In the present work we use it only as a diagnostic and do not discard solutions with $\chi^2>1$. For all isotopes we find solutions with $\chi^2\approx 1$, i.e. the transform is reproduced within the estimated many-body uncertainty.

\section{Validation}
\label{validation}

\subsection{Convergence of the integral transform} 
\begin{figure}[t]
   \centering
    \includegraphics[width=0.49\textwidth]{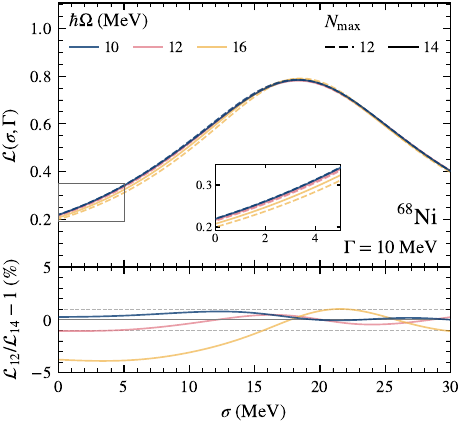}
    \caption{Upper panel: LIT of the $^{68}$Ni E1 response at $\Gamma=10$~MeV for $N_{\rm max}=12$ (dashed) and $N_{\rm max}=14$ (solid) and for $\hbar\Omega=10,12,16$~MeV; the inset enlarges the low-$\sigma$ region. Lower panel: relative difference between the $N_{\rm max}=12$ and $N_{\rm max}=14$ results for each $\hbar\Omega$.}
 \label{fig:68ni_convergence}
\end{figure}

We begin by studying the model-space convergence of response functions for the closed-shell $^{68}$Ni and $^{78}$Ni, for which model spaces up to $N_{\rm max}=14$ are accessible and which essentially bracket the mass range of the open-shell isotopes considered later. Fig.~\ref{fig:68ni_convergence} shows the LIT of the $^{68}$Ni E1 response at $\Gamma=10$~MeV for $N_{\rm max}=12$ and $14$ and for $\hbar\Omega=10,12,16$~MeV; the corresponding results for $^{78}$Ni are collected in the Appendix.
Even at $N_{\rm max}=14$ a small residual dependence on $\hbar\Omega$ remains. In the region $\sigma\lesssim 30$~MeV, however, the difference between the $N_{\rm max}=12$ and $14$ results is below the percent level for $\hbar\Omega=10$~MeV, and noticeably larger for the higher frequencies (lower panel of Fig.~\ref{fig:68ni_convergence}). Energies below 30 MeV are the most relevant for the dipole response, as they contain the GDR and possible low-lying pygmy strength, and it has been observed that smaller values of $\hbar\Omega$ are better suited to converge it (see e.g. Ref.~\cite{bonaiti2022,bonaiti2024}). Based on this analysis, we perform all calculations for the open-shell isotopes, for which we are limited to $N_{\rm max}=12$, with $\hbar\Omega=10$~MeV.

\subsection{Closed-shell nickel isotopes: $^{68}$Ni and $^{78}$Ni}
\begin{figure}[t]
   \centering
    \includegraphics[width=0.49\textwidth]{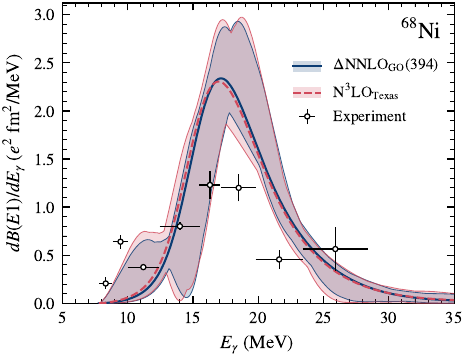}
    \caption{E1 strength distribution of $^{68}$Ni obtained with the $\Delta$NNLO$_{\rm GO}(394)$ and N$^3$LO$_\text{Texas}$
    interactions for $N_{\rm max}=14$, $\hbar\Omega=10$~MeV. The bands represent the inversion uncertainty. Experimental data from Ref.~\cite{rossi2013} include only the neutron-emission channel.}
 \label{fig:68ni_ham}
\end{figure}
Fig.~\ref{fig:68ni_ham} shows the E1 strength distribution of $^{68}$Ni, obtained by inverting the $N_{\rm max}=14$, $\hbar\Omega=10$~MeV transform for the $\Delta$NNLO$_{\rm GO}(394)$ and N$^3$LO$_\text{\rm Texas}$ interactions. The Hamiltonian dependence is much smaller than the inversion uncertainty, which for this nucleus is the largest among all isotopes considered: as mentioned in Sec.~\ref{inversion}, the coupled-cluster spectrum of $^{68}$Ni contains a discrete state just above the computed threshold, which the smooth basis functions of Eq.~(\ref{eq:basis}) can only accommodate with difficulty. The experimental data~\cite{rossi2013} contain only the one-neutron- and two-neutron-emission channels, whereas our calculation yields the inclusive photoabsorption strength, including proton emission. Therefore, above the proton threshold (15.431 MeV~\cite{ame2020}), we expect to overestimate the data. Up to that energy region, our result agrees with the data within uncertainties.
The results for $^{78}$Ni (see Appendix) are qualitatively similar, with a smaller inversion uncertainty and again a weak Hamiltonian dependence. In the following we therefore safely assume that the Hamiltonian dependence of the dipole response is subdominant with respect to the inversion uncertainty along the whole chain, and we employ only the $\Delta$NNLO$_{\rm GO}(394)$ interaction.

\subsection{$^{58}$Ni: photoabsorption and dipole polarizability}
\begin{figure}[t]
   \centering
    \includegraphics[width=0.49\textwidth]{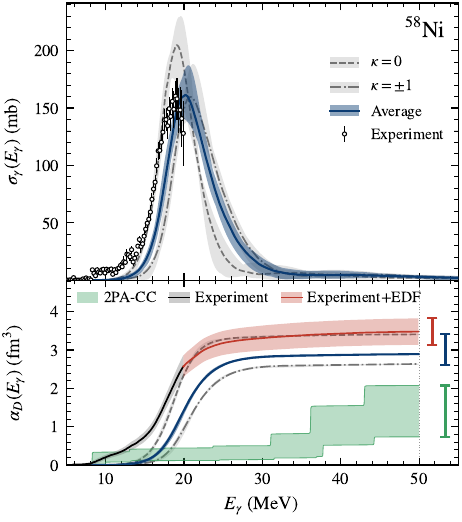}
    \caption{Upper panel: photoabsorption cross section of $^{58}$Ni for the $\kappa=0$ (dashed) and $\kappa=\pm1$ (dash-dotted) components of the dipole operator, Eq.~(\ref{eq:dipole_op}), and for their average, Eq.~(\ref{eq:average}) (solid); the bands represent the inversion uncertainty. Experimental data from Ref.~\cite{brandherm2025}. Lower panel: running sum of the dipole polarizability $\alpha_D$ for the same three cases, compared with the experimental extraction of Ref.~\cite{brandherm2025}, supplemented by a EDF calculation above 20~MeV, and with the two-particle-attached LIT-CC result of Ref.~\cite{brandherm2025}. The bars on the right indicate the saturated values with their uncertainties.}
 \label{fig:58ni_photoabs}
\end{figure}
The lightest open-shell isotope we consider is $^{58}$Ni, for which the photoabsorption cross section was recently measured~\cite{brandherm2025}. This nucleus therefore provides a direct test of the symmetry-broken framework. Since the reference state does not conserve angular momentum, we compute the response separately for the $\kappa=0$ and $\kappa=\pm1$ components of the dipole operator, Eq.~(\ref{eq:dipole_op}), and use their difference as a measure of the symmetry breaking (see Sec.~\ref{sec:lit}). The results are shown in the upper panel of Fig.~\ref{fig:58ni_photoabs}: the $\kappa=0$ (dashed) and $\kappa=\pm1$ (dash-dotted) components with their inversion uncertainties, and the averaged response of Eq.~(\ref{eq:average}) (solid line).
We find reasonable agreement with the data. The unprojected $\kappa=0$ component reproduces very well the low-energy slope of the GDR, while the averaged response is shifted by about $1.5$~MeV towards higher energies but recovers the strength of the GDR. The fragmentation of the low-lying strength at $8$--$12$~MeV seen in the experiment is not reproduced by our inversion.
We confirmed that the discretized response, calculated directly from the Lanczos eigenvalues and transition strengths, exhibits only negligible strength in this region compared with the GDR peak. We have also checked the sensitivity of the response to ground-state correlations by including triples excitations in the cluster operator (CCSDT-1)~\cite{watts1993}, while keeping the EOM operators at the 2p-2h level, in a smaller model space, $N_{\rm max}=8$, and found only a very small effect. 

In Ref.~\cite{brandherm2025}, also the dipole polarizability $\alpha_D$ was extracted from the data. Being proportional to the inverse-energy-weighted sum rule of the response, it is mostly sensitive to the low-energy part of the spectrum. As it is strongly correlated with symmetry energy parameters in the equation of state~\cite{roca-maza2013}, it has been the object of significant theoretical and experimental interest (see e.g. Refs.~\cite{birkhan2017,kaufmann2020,fearick2023}). %
The lower panel of Fig.~\ref{fig:58ni_photoabs} compares the experimental extraction, supplemented by an EDF calculation above 20~MeV, with the running sum of $\alpha_D$ obtained from our inverted responses, and from a previous coupled-cluster result obtained within the two-particle-attached (2PA-LIT-CC) framework~\cite{brandherm2025}. The latter underestimates $\alpha_D$ considerably above $10$ MeV, owing to an insufficient treatment of correlations in the $^{58}$Ni wavefunction, whereas the present result agrees with the experimental extraction within uncertainties. Our theoretical uncertainty on $\alpha_D$ is dominated by the lack of angular momentum projection, i.e. by the spread between the $\kappa=0$ and $\kappa=\pm1$ components. The inversion uncertainty on each component separately is negligible, reflecting the fact that all solutions of a given family exhaust the same inverse-energy-weighted sum rule.

\subsection{Neutron-rich open-shell isotopes}
\begin{figure*}[htb]
   \centering
    \includegraphics[width=\textwidth]{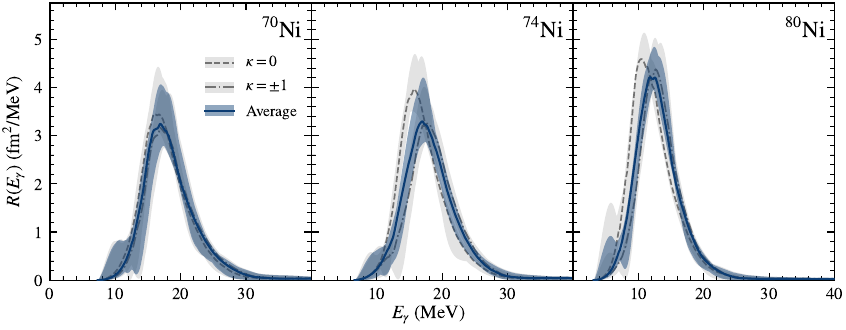}
    \caption{E1 response functions of $^{70}$Ni, $^{74}$Ni and $^{80}$Ni for the $\kappa=0$ (dashed) and $\kappa=1$ (dash-dotted) components of the dipole operator and for their average (solid), computed with the $\Delta$NNLO$_{\rm GO}(394)$ interaction for $N_{\rm max}=12$, $\hbar\Omega=10$~MeV. The bands represent the inversion uncertainty.}
 \label{fig:nickels70-74-80}
\end{figure*}

Fig.~\ref{fig:nickels70-74-80} shows the dipole response of $^{70}$Ni, $^{74}$Ni and $^{80}$Ni (the responses of $^{72,76,78}$Ni are qualitatively similar and are given in the Appendix). In all cases we predict low-lying pygmy strength, which for $^{80}$Ni peaks around 6~MeV. The $\kappa=0$ and $\kappa=\pm1$ components follow a systematic pattern: the $\kappa=0$ response peaks at lower energies and higher strength than the $\kappa=\pm1$ result. This behaviour was observed also in mean-field computations of the E1 response~\cite{arteaga2008,peru2008}. Averaging over the three components therefore reduces the peak strength and shifts the response towards higher energies with respect to $R_{\kappa=0}$, in qualitative agreement with what was found for the quadrupole response and its angular-momentum projection in Ref.~\cite{porro2024b}. The degree of symmetry breaking, measured by the difference between the $\kappa=0$ and $\kappa=\pm1$ results, varies along the chain: it is very small for $^{70}$Ni, more pronounced for $^{74}$Ni, and intermediate for $^{80}$Ni. This can be understood, at least in part, in terms of the filling of the neutron shell: $^{70}$Ni and $^{80}$Ni lie close to the closed-shell $^{68}$Ni and $^{78}$Ni, respectively, while $^{74}$Ni sits in the middle of the $1g_{9/2}$ shell.

\section{Astrophysical reaction rates}
\label{astro_rates}
Starting from the E1 response functions presented in the previous Section, we compute astrophysical neutron capture rates with the reaction code TALYS~\cite{koning2023}. Our focus is the impact of the E1 strength above the first particle threshold, i.e. the pygmy dipole region. To this end, we added to TALYS an option allowing for our $\gamma$-strength function to be supplied
as external input. Remaining settings are left at their default values. Among these, we list here some relevant choices. We use the Gilbert-Cameron model~\cite{gilbert1965} for the level density
and the global Koning-Delaroche nucleon-nucleus optical potential~\cite{koning2003}. We do not quantify the uncertainty associated with these inputs. For the level density, we note that the Gilbert-Cameron
model follows a trend similar to the experimental determination in
$^{70}$Ni~\cite{liddick2016}. A recent experiment~\cite{ronning2026} has attributed a M1 nature to the low-energy enhancement or upbend of the $\gamma$-strength function, which can significantly affect neutron capture~\cite{larsen2018}. We therefore retain the default
phenomenological M1 upbend provided by TALYS, together with a simple
modified Lorentzian (SMLO)~\cite{goriely2019} for the M1 spin-flip contribution. The default choice for width fluctuation corrections is set to the Moldauer model~\cite{moldauer1980}.

\begin{figure*}[htb]
   \centering
    \includegraphics[width=\textwidth]{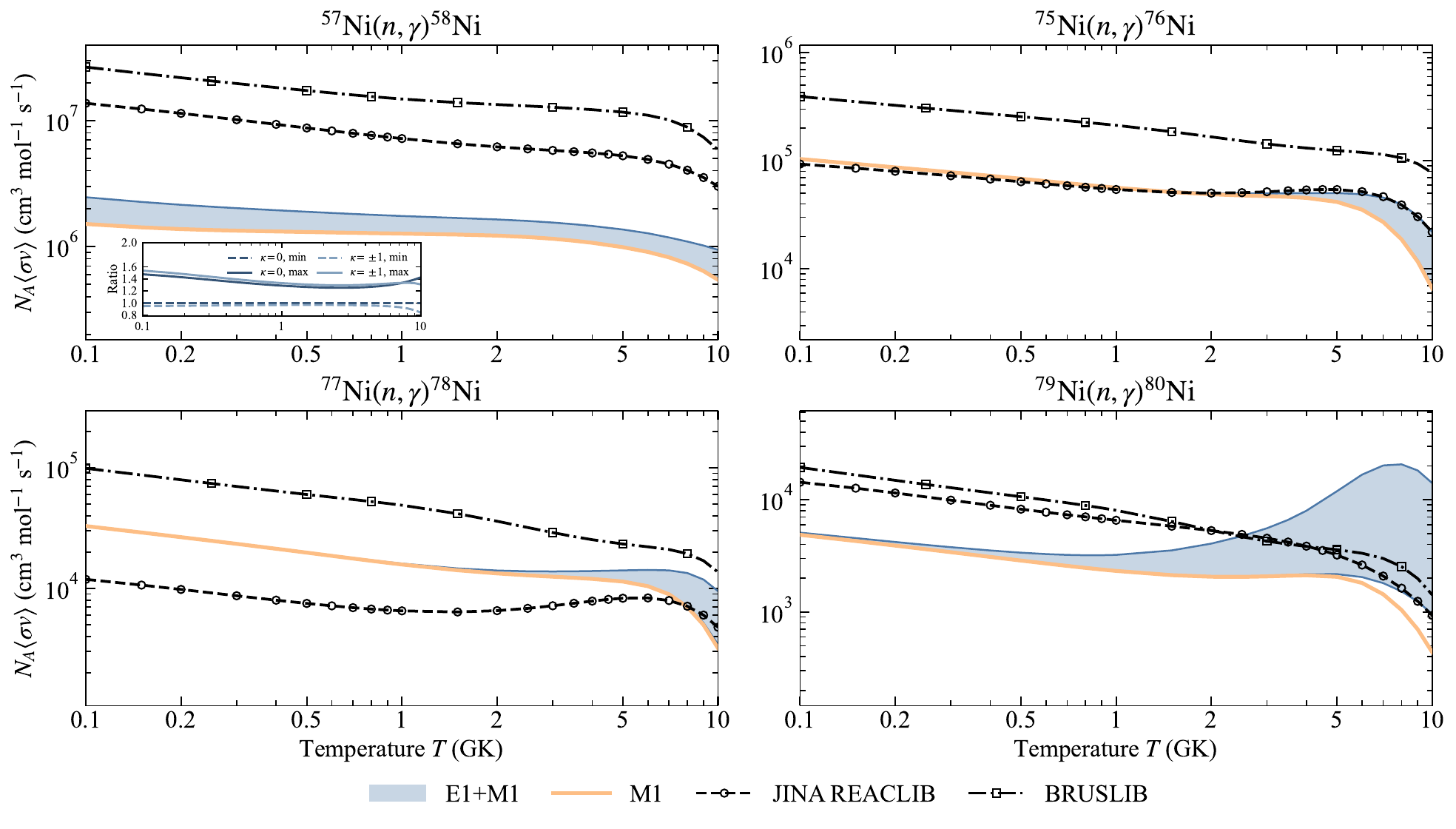}
    \caption{Astrophysical reaction rates for $^{57}$Ni$(n,\gamma)^{58}$Ni, $^{75}$Ni$(n,\gamma)^{76}$Ni, $^{77}$Ni$(n,\gamma)^{78}$Ni and $^{79}$Ni$(n,\gamma)^{80}$Ni as a function of temperature $T$ obtained from TALYS employing our coupled-cluster result for the E1 strength and TALYS  default phenomenological model for the M1 strength (blue band). The result obtained using only the M1 phenomenological model for the $\gamma$-strength function is shown in orange. JINA REACLIB~\cite{JINA_REACLIB} and BRUSLIB~\cite{xu2013} reaction rates are also reported. See text for details. }
 \label{fig:rates}
\end{figure*}

Fig.~\ref{fig:rates} shows the astrophysical reaction rates $N_A\braket{\sigma v}$  we obtained for the $^{57}$Ni$(n,\gamma)^{58}$Ni, $^{75}$Ni$(n,\gamma)^{76}$Ni, $^{77}$Ni$(n,\gamma)^{78}$Ni and $^{79}$Ni$(n,\gamma)^{80}$Ni capture reactions. Neutron capture rates on $^{67,69,71,73}$Ni are shown in the Appendix. We compare two calculations: one
including only the phenomenological M1 strength (upbend and spin-flip with no E1 contribution), shown in orange, and one including in addition our E1 strength
above the first particle threshold (shown as a blue band encompassing our theoretical uncertainty). For $^{77}$Ni$(n,\gamma)^{78}$Ni,
which forms the closed-shell $^{78}$Ni as compound nucleus, the band is
obtained from the maximum and minimum E1 curves arising from the inversion. If the compound nucleus is instead open-shell, the band is defined by the maximum and minimum curves over both $\kappa=0$ and $\kappa=\pm1$, thus accounting for both the inversion and symmetry-breaking effects. 
The lower edges of the ``E1+M1'' bands in Fig.~\ref{fig:rates} correspond to inversions which have the least low-lying strength. They overlap with the M1-only result in all cases except $^{79}$Ni$(n,\gamma)^{80}$Ni between 5 and 10~GK. The upper edge, conversely, corresponds to inversions which predict a large pygmy strength.
The band therefore spans from a negligible E1 contribution to the largest effect compatible with
our E1 response.

The uncertainty on the rate associated with the E1 response is dominated
by the inversion. This is illustrated in the inset of Fig.~\ref{fig:rates} for $^{57}$Ni$(n,\gamma)^{58}$Ni, where the maximum and minimum edges for $\kappa=0$ and $\kappa=\pm1$ are shown normalized to the minimum edge for $\kappa=0$. The spread between the maximum and minimum for a given $\kappa$ is much larger than the difference between $\kappa=0$ and $\kappa=\pm1$. Small discrepancies between the two appear only at high temperatures, consistent with the fact that the GDR, which dominates the E1 response of $^{58}$Ni, peaks at higher energies for $\kappa=\pm1$, as shown in Fig.~\ref{fig:58ni_photoabs}.

The relative contribution of the E1 strength increases with temperature and neutron number. Its temperature dependence can be understood within the compound-nucleus picture. For the neutron-rich isotopes considered here, the first particle-emission threshold is the neutron separation energy $S_n$. Capture of a neutron with center-of-mass energy $E_n$ on a target in its ground state produces a compound nucleus at excitation energy $S_n+E_n$. A subsequent transition to a state with excitation energy $E_x$ emits a photon with energy $E_\gamma=S_n+E_n-E_x$. Consequently, the E1 strength above $S_n$ contributes only to transitions feeding states with $E_x< E_n$. As the temperature increases, the neutron energy distribution samples higher energies, allowing more final states to be reached through such transitions and a wider interval of the above-threshold E1 strength to contribute. The relevant thermal energy scale is of the order of MeV at 10~GK.

For capture forming the even-even isotopes from $^{68}$Ni to $^{78}$Ni, the calculated E1 contribution becomes appreciable mainly at temperatures of 3--10~GK, and its relative impact increases with neutron number. At the upper edge of the uncertainty band, the enhancement over the M1-only baseline at 5~GK increases from $16\%$ for $^{67}$Ni$(n,\gamma)^{68}$Ni to $24\%$ for $^{77}$Ni$(n,\gamma)^{78}$Ni. At 10~GK, the corresponding total rates reach 1.8 and 3 times their respective M1-only baselines. 

The increase with neutron number follows the growth of the low-lying E1 strength above $S_n$. For $^{79}$Ni$(n,\gamma)^{80}$Ni, the E1 strength affects the rate already at lower temperatures. This reflects the large pygmy dipole mode of $^{80}$Ni close to threshold, shown in Fig.~\ref{fig:nickels70-74-80}. $^{57}$Ni$(n,\gamma)^{58}$Ni, instead, is a special case. There, the first
particle threshold of the compound nucleus $^{58}$Ni is the proton
separation energy $S_p$ ($8.17$ MeV~\cite{ame2020}), which lies several MeV below $S_n$ ($12.22$ MeV~\cite{ame2020}). Our E1
strength is therefore probed by transitions to all final states with $E_x < S_n + E_n - S_p$. This explains the sensitivity of the rate to the E1 strength in the whole temperature range under consideration.

A similar analysis was carried out in Ref.~\cite{ghosh2026} within EDF theory. There, the impact of the PDR is quantified by the ratio $N_A\braket{\sigma v}_{\mathrm{PDR+GDR}}/N_A\braket{\sigma v}_{\mathrm{GDR}}$ between the rates calculated with and without the PDR, with the GDR being retained in both. They find a ratio between 1 and 2 for $^{58}$Ni to $^{76}$Ni at 1--10~GK, and almost 10 for $^{80}$Ni at 1~GK. In our case, we compare the full E1 contribution above threshold with respect to a calculation without E1 strength. Our results are therefore not directly comparable to Ref.~\cite{ghosh2026}. Still, we find the same trend of an increasing E1 effect with neutron number. 

Finally, Fig.~\ref{fig:rates} compares our rates with the tabulated rates
of JINA REACLIB~\cite{JINA_REACLIB} and BRUSLIB~\cite{xu2013}. Both libraries are based on Hauser-Feshbach calculations, but with different nuclear inputs. The two libraries show similar trends, with BRUSLIB predicting higher reaction rates with respect to JINA REACLIB in most cases. Our results are consistent with the two libraries within one order of magnitude,
which is comparable to typical uncertainties of statistical model rates
in this region~\cite{liddick2016}. This qualitative agreement is encouraging, but the
comparison should be considered as indicative, since none of the three
sets of rates includes a full uncertainty quantification.

\section{Conclusions}
\label{conclusions}
We computed the electric dipole response of nickel isotopes up to $^{80}$Ni with the LIT-CC method and interfaced the obtained ab initio E1 strength with Hauser--Feshbach calculations of astrophysical neutron capture rates with the reaction code TALYS. 

To reach mid-shell isotopes, we developed a formulation of the LIT-CC method based on an axially symmetric reference state and benchmarked it against experimental data for $^{58}$Ni and $^{68}$Ni. In this framework, the spread between the $\kappa=0$ and $\kappa=\pm1$ components of the dipole response measures the breaking of rotational symmetry. Along the isotopic chain, the $\kappa=\pm1$ components systematically yield lower strength and higher GDR energies than the $\kappa=0$ component, with differences that decrease near shell closures. While restoring rotational invariance would require angular momentum projection, the astrophysical rates, turn out to be almost insensitive to this effect. For all even-even isotopes from $^{68}$Ni to $^{80}$Ni we find low-lying E1 strength above the neutron threshold, whose magnitude is at present affected by a sizable uncertainty stemming from the inversion of the integral transform.

We assessed the impact of the computed E1 strength by comparing neutron capture rates obtained with a phenomenological M1 strength alone with those including both E1 and M1 contributions. For capture forming the even-even isotopes from $^{68}$Ni to $^{78}$Ni, the E1 strength affects the rates at temperatures between $3$ and $10$~GK, and the relative enhancement over the M1-only result grows with neutron number. For $^{79}$Ni$(n,\gamma)^{80}$Ni, deviations from the M1-only result appear already at $0.5$~GK, owing to the large pygmy dipole strength of $^{80}$Ni close to threshold.

This work opens two directions for future development. On the nuclear structure side, the reliance on phenomenological input could be reduced further by extending the \textit{ab initio} description to the E1 and M1 strength below the first particle-emission threshold. The M1 contribution is of particular interest, as it drives the low-energy upbend of the $\gamma$-ray strength function~\cite{ronning2026}. This requires angular-momentum projection in the calculation of transition strengths, which we plan to investigate. At the same time, since the uncertainty of the rates is dominated by the inversion of the integral transform, an improved inversion procedure, in particular at low energies, would reduce it substantially. On the astrophysical side, the rates obtained here can be used directly in nucleosynthesis simulations. It would be interesting to see whether the E1 effects found in this work, in particular the low-temperature enhancement for $^{80}$Ni, leave an imprint on the abundances below $A\approx100$.

\section*{Acknowledgements}
We thank Gaute Hagen for giving us access to the ORNL coupled-cluster code and Gaute Hagen and Thomas Papenbrock for useful discussions. F.B. thanks the Chalmers Nuclear Theory group for their hospitality and support during her two research visits in October 2025 and September-October 2026. Financial support for the second visit was provided by the Chalmers University of Technology Jubilee Professorship awarded to Gaute Hagen. J.E.S. thanks the ORNL Nuclear Theory group for their hospitality and support during her research visit in February 2025. We would like to express special thanks for their hospitality and support to the Mainz Institute for Theoretical Physics (MITP) of the Cluster of Excellence PRISMA+ (Project ID 390831469), Institute for Nuclear Theory at the University of Washington and ECT* for support at the workshop “Responses of strongly interacting systems from femto to astrophysical scales” during which this work has been developed. This research was supported in part by the INT's U.S. Department of Energy grant No. DE-FG02- 00ER41132. This work was supported partially through EUSTIPEN (Europe-U.S. Theory Institute for Physics with Exotic Nuclei), which is supported by FRIB Theory Alliance under DOE grant number DE-SC0013617. This work was supported by the U.S. Department of Energy, Office of Science, Office of Nuclear Physics, under the FRIB Theory Alliance award DE-SC0013617 and award No.~DE-FG02-96ER40963; by the U.S. Department of Energy, Office of Science, Office of Advanced Scientific Computing Research and Office of Nuclear Physics, Scientific Discovery through Advanced Computing (SciDAC) program (SciDAC-5 NUCLEI). This research used resources of the Oak Ridge Leadership Computing Facility located at Oak Ridge National Laboratory, which is supported by the Office of Science of the Department of Energy under contract No. DE-AC05-00OR22725. Computer time was provided by the Innovative and Novel Computational Impact on Theory and Experiment (INCITE) program and by by the Institute for Cyber-Enabled Research at Michigan State University. \\


\appendix
\section{Model-space convergence and E1 responses of \texorpdfstring{$^{72,76,78}$Ni}{72,76,78Ni}}
\label{app:responses}

\begin{figure}[hbt]
   \centering
    \includegraphics[width=0.49\textwidth]{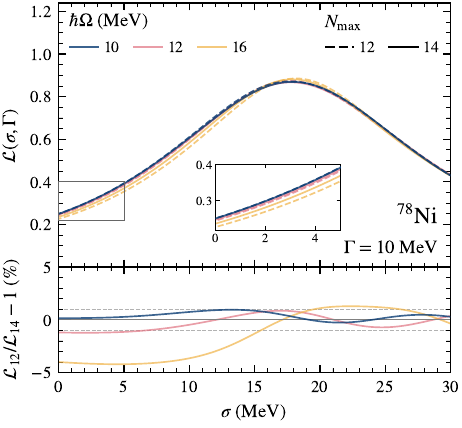}
    \caption{Same as Fig.~\ref{fig:68ni_convergence}, for $^{78}$Ni.}
 \label{fig:78ni_convergence}
\end{figure}

Fig.~\ref{fig:78ni_convergence} shows the model-space convergence of the LIT for $^{78}$Ni. The pattern closely resembles that of $^{68}$Ni (Fig.~\ref{fig:68ni_convergence}). For $\hbar\Omega=10$~MeV, the $N_{\rm max}=12$ and $N_{\rm max}=14$ transforms at $\Gamma=10$~MeV differ by less than 1\% for $\sigma\lesssim 30$~MeV, which is the region relevant for the dipole response.

\begin{figure*}[htb]
   \centering
    \includegraphics[width=0.98\textwidth]{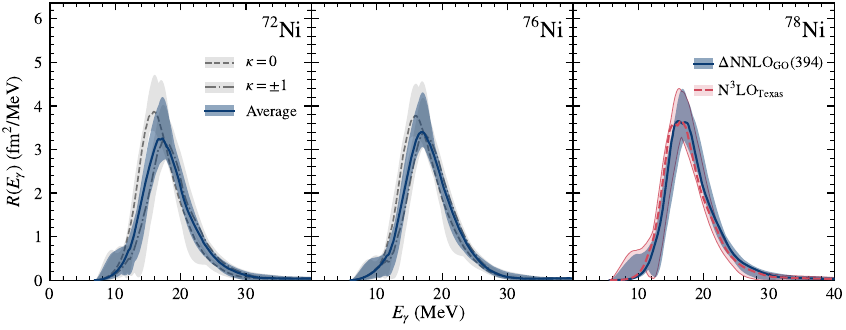}
    \caption{Same as Fig.~\ref{fig:nickels70-74-80}, for $^{72}$Ni, $^{76}$Ni and $^{78}$Ni. For the closed-shell $^{78}$Ni, the results obtained with the $\Delta$NNLO$_{\rm GO}(394)$ and N$^3$LO$_{\rm Texas}$ interactions are compared.}
 \label{fig:responses_72_76_78}
\end{figure*}

Fig.~\ref{fig:responses_72_76_78} shows the E1 responses of $^{72,76,78}$Ni, which were not shown in Sec.~\ref{validation}. Together with Figs.~\ref{fig:58ni_photoabs} and~\ref{fig:nickels70-74-80}, they complete the set of inputs for the reaction-rate calculations of Sec.~\ref{astro_rates}. For $^{78}$Ni we also show the result obtained with the N$^3$LO$_{\rm Texas}$ interaction, which gives an estimate of the Hamiltonian dependence. As in the isotopes discussed in the main text, all three nuclei show low-lying pygmy strength. The size of the symmetry breaking follows the pattern found in Sec.~\ref{validation}. It is comparable in $^{72}$Ni and $^{74}$Ni, which lie near the middle of the $1g_{9/2}$ shell, and it is smaller in $^{76}$Ni, which, like $^{70}$Ni, lies close to a shell closure.

\section{Astrophysical reaction rates for neutron capture on $^{67,69,71,73}$Ni}
\begin{figure*}[htb]
   \centering
    \includegraphics[width=\textwidth]{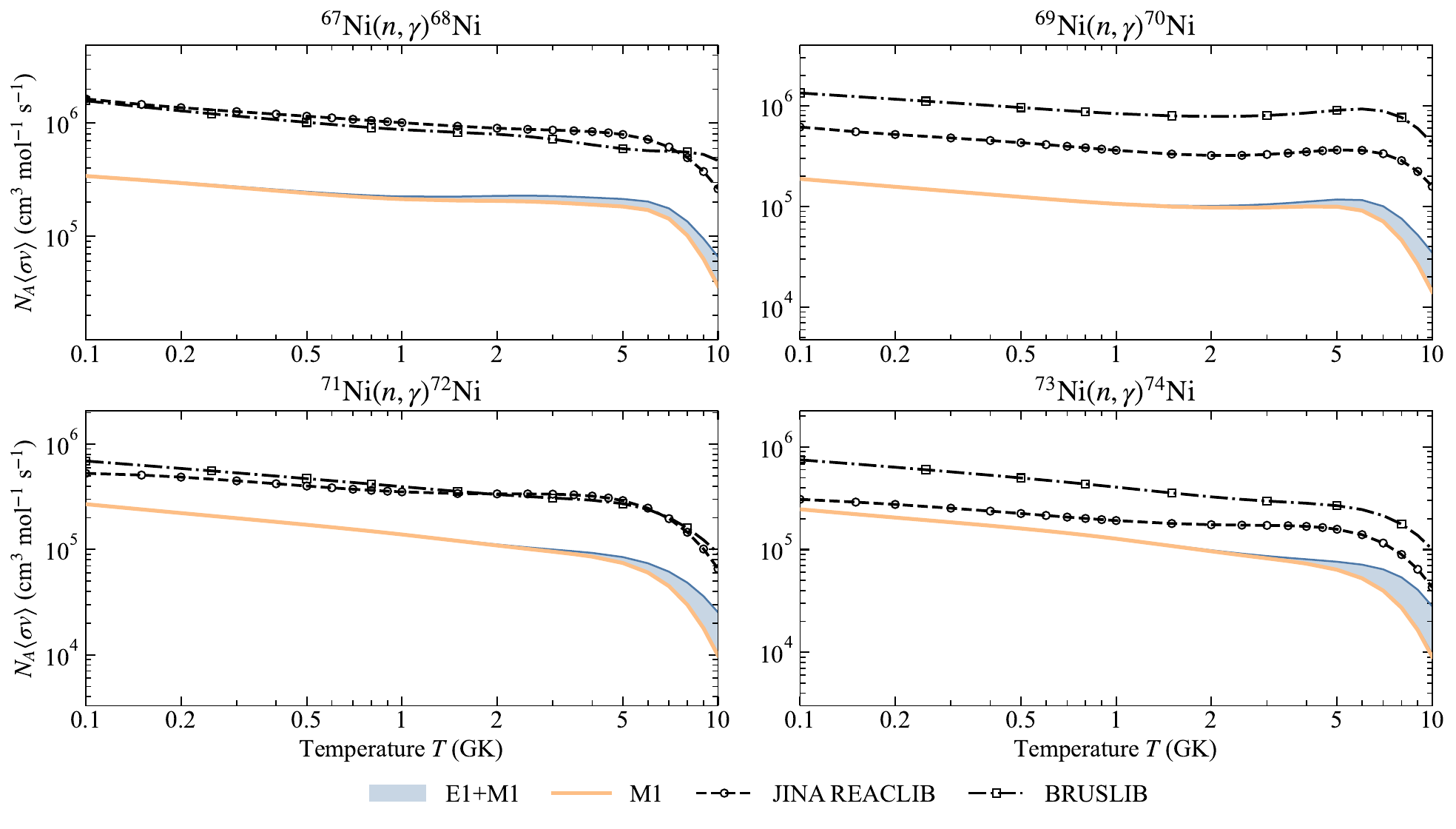}
    \caption{Same as Fig.~\ref{fig:rates}, for $^{67}$Ni$(n,\gamma)^{68}$Ni, $^{69}$Ni$(n,\gamma)^{70}$Ni, $^{71}$Ni$(n,\gamma)^{72}$Ni and $^{73}$Ni$(n,\gamma)^{74}$Ni. }
 \label{app:remaining_rates}
\end{figure*}

In Fig.~\ref{app:remaining_rates}, we show the astrophysical reaction rates for neutron capture on $^{67,69,71,73}$Ni, which were not shown in Sec.~\ref{astro_rates}. 

In analogy to what was observed in Fig.~\ref{fig:rates} for neutron capture on $^{75,77}$Ni, low-lying E1 strength enhances the rates mainly at high temperatures, between 3 and 10 GK. Our results are consistent with JINA REACLIB and BRUSLIB within one order of magnitude. 

\clearpage
\onecolumn

\bibliographystyle{elsarticle-num}
\bibliography{master}

\end{document}